\documentstyle[preprint,aps,tighten]{revtex}

\begin{document}


\def \GeV{{\rm \enspace GeV}}
\def \TeV{{\rm \enspace TeV}}
\def \beq{\begin{equation}}
\def \eeq{\end{equation}}
\def \beqa{\begin{eqnarray}}
\def \eeqa{\end{eqnarray}}
\def \up{\uparrow}
\def \down{\downarrow}
\def \ts{\thinspace}
\def \nts{\negthinspace}
\def \ns{\enspace}
\def \twothree{$2\rightarrow 3$}
\def \twotwo{$2\rightarrow 2$}
                
\draft
\preprint{
  \parbox{2in}{Fermilab--Pub--99/361-T \\
  McGill/99--39 \\
  SLAC-PUB-8317 \\
  hep-ph/9912458
}  }

\title{Single Top Quark Production at the LHC: Understanding Spin}
\author{Gregory Mahlon  \cite{GDMemail}}
\address{Department of Physics, McGill University, \\
3600 University St., Montr\'eal, Qu\'ebec  H3A 2T8, Canada }
\author{Stephen Parke \cite{SPemail}}
\address{Theoretical Physics Department\\
Fermi National Accelerator Laboratory \\
P.O. Box 500, Batavia, Illinois  60510, U.S.A. \\
{\rm and} \\
Theory Group\\
Stanford Linear Accelerator Center\\
Stanford, California 94305, U.S.A.}
\date{December 22, 1999}
\maketitle
\begin{abstract}
We show that the single top quarks produced in the $Wg$-fusion 
channel at a proton-proton collider at a center-of-mass energy 
$\sqrt{s}=14\TeV$ posses a high degree of polarization in terms 
of a spin basis which decomposes the top quark spin in its rest 
frame along the direction of the spectator jet.  
A second useful spin basis is the $\eta$-beamline basis, which 
decomposes the top quark spin along one of the two beam directions, 
depending on which hemisphere contains the spectator jet.
We elucidate the interplay between the two- and three-body final 
states contributing to this production cross section in the context 
of determining the spin decomposition of the top quarks,
and argue that the zero momentum frame helicity is undefined.  
We show that the usefulness of the spectator and $\eta$-beamline
spin bases is not adversely affected by the cuts required to separate 
the $Wg$-fusion signal from the background.
\end{abstract}


One of the many physics goals of the CERN Large Hadron Collider
(LHC) program is a detailed study of the top quark.  
With a measured mass of 173.8 $\pm$ 5.2 GeV\cite{PDG1999},
the top quark is by far the heaviest known fermion, and the
only known fermion with a mass at the electroweak symmetry-breaking
scale.  Thus, it is hoped that a detailed study of how the
top quark couples to other particles will be of great utility 
in determining if the Standard Model mechanism for electroweak
symmetry-breaking is the correct one, or if some type of new
physics is responsible.
Angular correlations among the decay products of polarized top
quarks provide a useful handle on these couplings.
One consequence of the large top quark mass is that the 
time scale for the top quark decay, set by its decay width
$\Gamma_t$, is much shorter than the typical time 
required for QCD interactions to randomize its spin\cite{Bigi}:
a top quark produced with spin up decays as a top quark 
with spin up.  The Standard Model
$V{-}A$ coupling of the $W$ boson to the top quark leaves an
imprint in the form of strong angular correlations among
the decay products of the top quark\cite{Spin}.  

The purpose of this letter is to 
demonstrate that single top quark production in the $Wg$ fusion 
channel at LHC energies provides a copious source of 
polarized top quarks.
Although possessing a larger production cross section,
top quark {\it pairs}\ at the LHC do not dominantly
populate a single spin configuration in any basis,
because the initial state is primarily $gg$\cite{TopPairsHadronic}.
On the other hand,
the $Wg$ fusion channel is the largest source of single top
quarks at the LHC.  At the most basic level, $Wg$ fusion is
an electroweak process, with the produced top quarks coupled
directly to a $W$ boson.  Therefore, it is not surprising to learn
that these top quarks are strongly polarized.
However, as has been shown in studies for other 
colliders\cite{TopPairsHadronic,TopPairsLeptonic,TopPairsOptimal,TeV1top}, 
the appropriate spin basis for the top quark is not the
traditional helicity basis.  The essential point is that unless the
particle whose spin is being studied is produced in the 
ultrarelativistic regime, there is no reason to believe
that the helicity basis will provide the simplest description
of the physics involved.  Top quarks produced in $pp$ collisions
via $Wg$ fusion at a center of mass energy $\sqrt{s}=14\TeV$
typically posses a speed of only
$\beta\sim 0.6$ in the zero momentum frame
(ZMF).  
Furthermore, the helicity of a massive particle is frame-dependent:
the direction of motion of the top quark changes as we boost from
frame to frame.  
This is significant, since, as we shall show, it is not possible
to unambiguously define the ZMF.  Thus, although we can pin
down the ZMF well enough to say that the typical speed of the
top quarks is $\beta\sim 0.6$ in that frame, we cannot do so
with the precision required to compute the top quark spin
decomposition in the ZMF helicity basis.  
Instead, we are left with the options of measuring the top quark
helicity in the laboratory frame (LAB helicity basis),
or using some other basis.
Fortunately, it is simple to construct a spin basis in which
well over 90\% of the top quarks are produced in one of the two
possible spin states.

\begin{figure}[h]

\vspace*{6cm}
\includegraphics{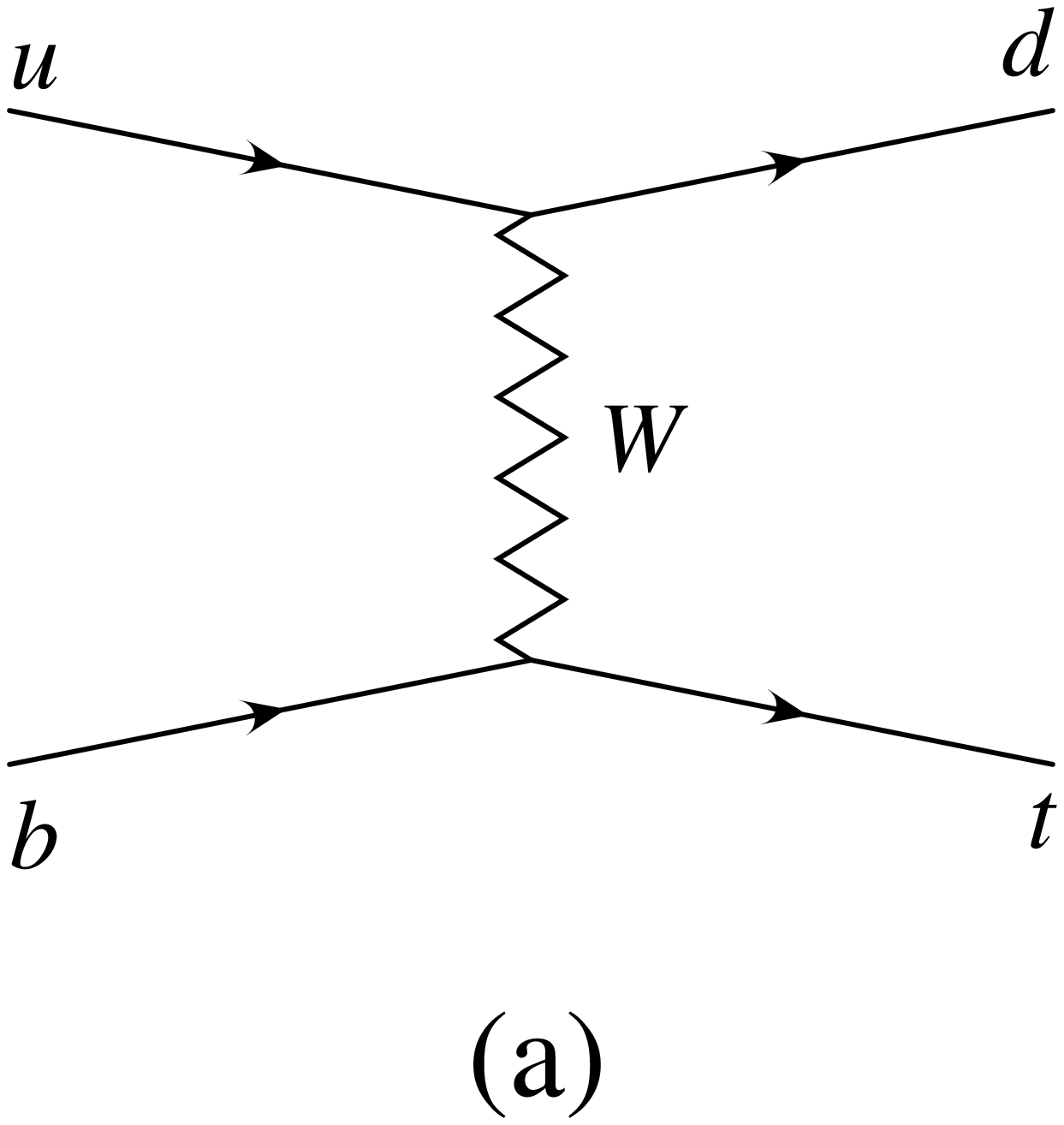}
\includegraphics{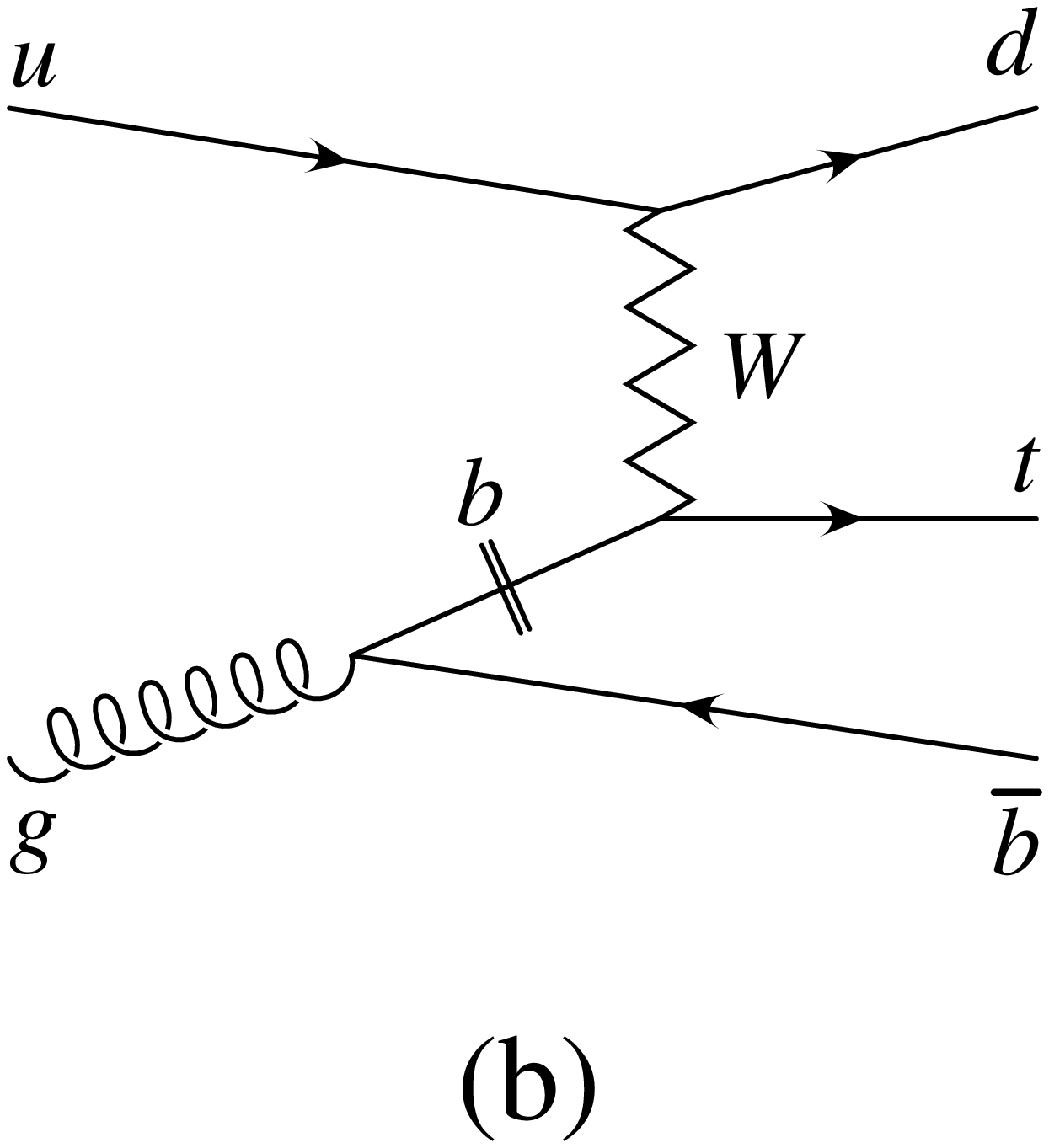}
\includegraphics{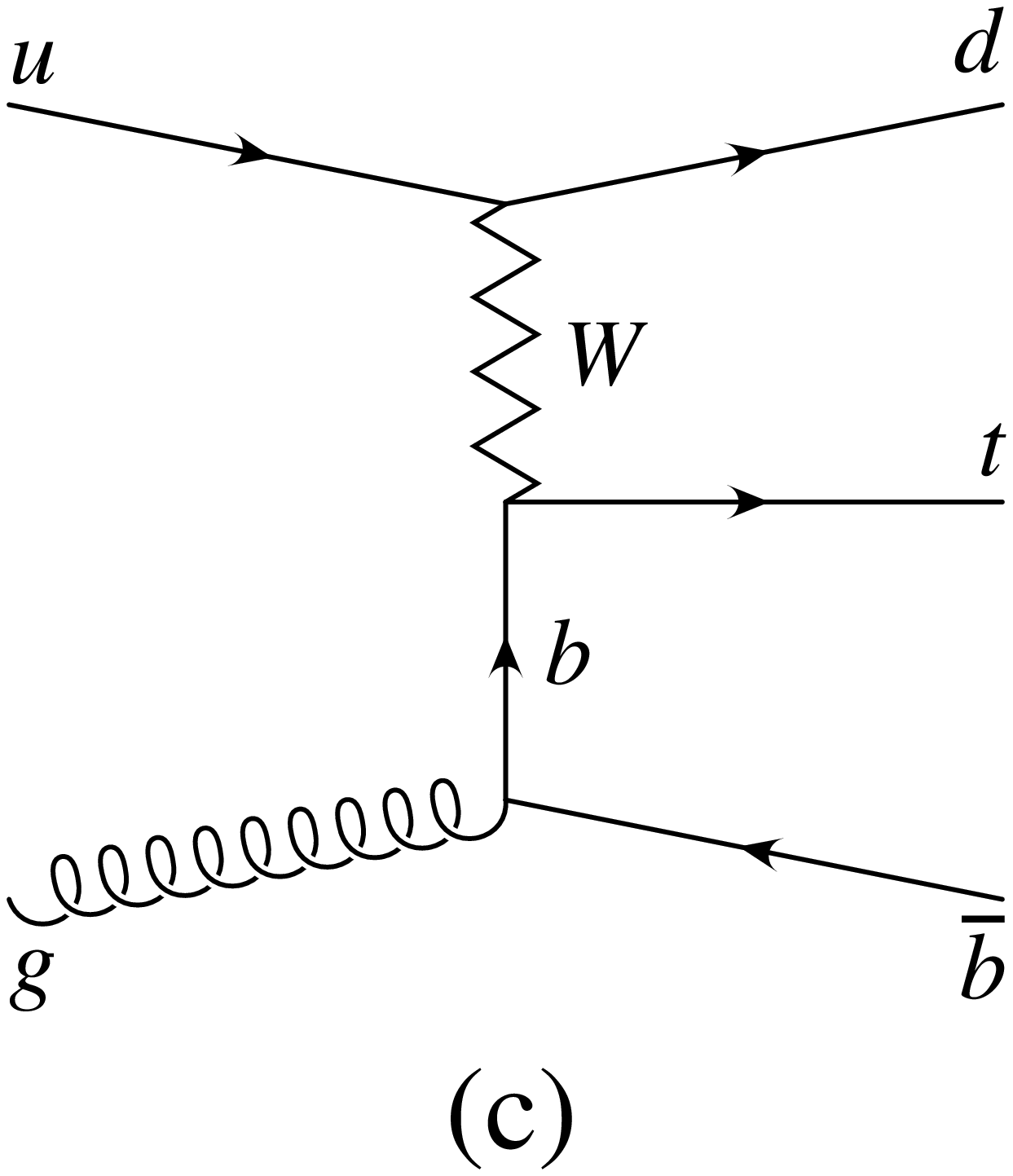}
\includegraphics{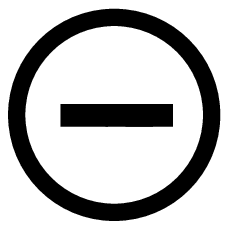}
\includegraphics{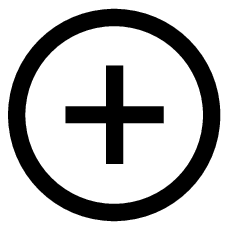}
\vspace{1.0cm}

\caption[]{Representative Feynman diagrams for 
single top quark production via $Wg$ fusion. 
}
\label{WgFusionDiagrams}
\end{figure}

We begin by outlining
the computation of the single 
top quark production cross section, which is shown schematically
in Fig.~\ref{WgFusionDiagrams}.
The calculation which we will use for our spin analysis 
may be described as ``leading order plus resummed large logs.''
For simplicity,
we do not include the additional tree-level \twothree\
and one-loop \twotwo\ diagrams which would be required for a
full next-to-leading order computation.
The neglected contributions turn
out to be numerically small (about 2.5\% of the total) at LHC 
energies\cite{Stelzer1}.

Early calculations of the $Wg$ fusion process were based solely
on the \twothree\ diagrams of 
Fig.~\ref{WgFusionDiagrams}c~\cite{Dawson,WillenbrockDicus,%
DawsonWillenbrock,Yuan1,EllisParke,UNK,CarlsonPLB}.
These diagrams are dominated by the
configuration where the final state $\bar{b}$ quark is nearly
collinear with the incoming gluon.  In fact, they become singular
as the mass of the $b$ quark is taken to zero.
This mass singularity appears as the large logarithm
$\ln(m_t^2/m_b^2)$.\footnote{More precisely, this logarithm reads
$\ln[(Q^2+m_t^2)/m_b^2]$ where $Q^2$ is the virtuality of the
$W$ boson.}
Furthermore, at each order in the strong coupling, there are
logarithmically enhanced contributions, converting the
perturbation expansion from a series in $\alpha_s$ to
one in $\alpha_s \ln(m_t^2/m_b^2)$.  To deal with this situation,
a formalism which sums these collinear logarithms to all orders
by introducing a $b$ quark parton distribution function
has been developed~\cite{HQpdf1,HQpdf2,HQpdf3}
and subsequently applied to $Wg$ 
fusion~\cite{Stelzer1,Bordes1,Bordes2,Bordes3}.
The large logarithms which caused the original perturbation
expansion to converge slowly are resummed to all orders and
absorbed into the $b$ quark distribution, which turns out
to be perturbatively calculable. 
Once the $b$ quark distribution has been introduced, we must
reorder perturbation theory, and begin with the \twotwo\ process
shown in Fig.~\ref{WgFusionDiagrams}a.  
The \twothree\ process then becomes a
correction to the \twotwo\ contribution.  
However, because the logarithmically enhanced terms within
the \twothree\ contribution have been
summed into the $b$ quark distribution, there is overlap between
the \twotwo\ and \twothree\ processes:  simply summing their
contributions will result in overcounting.
To account for this, we should subtract that portion of the \twothree\
diagram where the gluon splits into a (nearly) collinear
$b\bar{b}$ pair.  Schematically, we indicate this by the diagram
in Fig.~\ref{WgFusionDiagrams}b.
Equivalently, we should subtract the first
term from the series of collinear logarithms which were summed
to produce the $b$ quark distribution.  This point of view
is reflected by the prescription for computing 
Fig.~\ref{WgFusionDiagrams}b:
we simply use the \twotwo\ amplitude, but we replace the 
$b$ quark parton distribution function with the (lowest-order)
probability for a gluon to split into a $b\bar{b}$ pair:
\beq
b_0(x,\mu^2) =
{{\alpha_s(\mu^2)}\over{2\pi}}
\ln\biggl({{\mu^2}\over{m_b^2}}\biggr)
\int_x^1 
{{dz}\over{z}}
P_{qg}(z) g\biggl({{x}\over{z}},\mu^2\biggr).
\label{b-overlap}
\eeq
Eq.~(\ref{b-overlap}) contains the DGLAP splitting function
\beq
P_{qg}(z) = \hbox{$1\over2$}[ z^2 + (1-z)^2 ].
\eeq
The total single top quark production cross section then consists
of the \twotwo\ process minus the overlap plus the \twothree\ process.
As is emphasized in Ref.~\cite{HQpdf3}, the division among the
three kinds of contributions is arbitrary and depends upon
our choice of the QCD factorization scale.  Different choices
in factorization scale correspond to a reshuffling of the
contributions among the three terms.  

The production cross 
sections for single $t$ and $\bar{t}$ quarks will be unequal
at the LHC.
An initial state $u$, $\bar{d}$, $\bar{s}$, or $c$ quark is
required for $t$ production, whereas $\bar{t}$ production
requires an initial state $\bar{u}$, $d$, $s$ or $\bar{c}$ quark.
Since the LHC is a $pp$ collider, we expect more $t$ quarks
than $\bar{t}$ quarks, since the protons contain more valence
$u$ quarks than $d$ quarks.  This expectation is met 
by the total cross sections we obtain:  159 pb for $t$ production
and 96 pb for $\bar{t}$ production.\footnote{All of the cross sections
reported in this paper were computed using the CTEQ5HQ parton
distribution functions\protect\cite{CTEQ5}, two-loop running $\alpha_s$,
and the factorization scales advocated in Ref.~\protect\cite{Stelzer1}.} 
Table~\ref{Flavors}
summarizes the contributions from each flavor of light quark
in the initial state.
We see that for $t$ production, the initial state contains an
up-type quark 80\% of the time, while for $\bar{t}$ production,
the initial state contains a down-type quark 69\% of the time.
In the following discussion, we will talk in terms of the
dominant initial states, although when presenting the final
spin beakdowns, all flavors will be included.


We are now ready to discuss the spin of the top quarks produced
at LHC energies, beginning with the \twotwo\ contributions. 
For a final state $t$, the dominant \twotwo\ process is 
$ub\rightarrow dt$.  In the ZMF of the initial state partons, 
the outgoing $t$ and $d$ quarks are back-to-back.
Now the initial state contains a massless $u$ quark and an
effectively massless $b$ quark.  Since they couple to a $W$ boson,
we know they have left-handed chirality.  Since they are both
ultrarelativistic fermions, this left-handed chirality translates
into left-handed helicity.  Thus, the initial spin projection is
zero.  The final state $d$ quark is also massless, and so its
left-handed chirality also implies left-handed helicity.  Conservation
of angular momentum then leads to the $t$ quark having left-handed
helicity {\it in this frame}.  Since the $t$ quark is massive, boosting
to another frame will, in general, introduce a right-handed helicity
component.   In particular, if we measure the helicity of
the top quark in the laboratory frame instead of the ZMF, we find 
that it is left-handed only 66\% of the time.   

Turning to the \twothree\ process, we find that the addition of a
third particle to the final state frees the top quark from its
obligation to have left-handed helicity in the ZMF.  In fact,
we find that left-handed tops are produced only 82\% of the time
by this process.  Again, this number changes if we boost out
of the ZMF:  the fraction of left-handed helicity tops is only 
59\% in the lab frame.

When we come to the overlap contribution, we discover that it
is not possible to unambiguously define the ZMF.
Should we define the ZMF in terms of the light quark 
and gluon, or in terms of the
light quark and the $b$ quark which descended from the gluon via
splitting?  These two frames are different, 
and, as we have already argued,
the helicity of the top quark is not invariant under the longitudinal
boosts connecting these two frames.

Another way of illustrating the difficulty is to consider the
question of experimentally reconstructing the ZMF.  To
determine the ZMF, we would have to account for {\it all}\
of the final state particles.  With real detectors, this is
clearly impossible, since particles with small transverse
momenta
or very large pseudorapidity tend to be missed.
But the \twothree\ process frequently contains a low-$p_T$ $\bar{b}$
quark, which is likely to escape detection.
Even with a perfect detector, it is still not possible to
decide whether a given event came from the \twotwo\
diagram of Fig.~\ref{WgFusionDiagrams}a or the \twothree\
diagram of Fig.~\ref{WgFusionDiagrams}c.  
The point is, a perfect
detector would also track the proton remnants as well as the actual
scattering products.  Since there is no intrinsic bottom in the
proton, after a $2\rightarrow2$ interaction there would be a $\bar{b}$
quark among the proton remnants hitting our ``perfect'' detector.
As far as the detector is concerned, such a $\bar{b}$ quark would
look identical to the $\bar{b}$ quark generated by the \twothree\
process.  The best that could be done is to observe that a
$\bar{b}$ quark associated with the proton remnant would tend to
have a much smaller $p_T$ than one associated with the \twothree\
diagram.  However, the kinematics of the two processes overlap,
rendering the location of the dividing line arbitrary. 
This is precisely the physics of the statement made earlier
that the division of contributions among the \twotwo, \twothree,
and overlap terms is arbitrary, and depends on the QCD factorization
scale\cite{HQpdf3}.

Rather than use the (undefined) ZMF helicity basis, we should
decompose the top quark spin in a manner which does not depend
on any particular frame.  The spectator basis, introduced in
Ref.~\cite{TeV1top}, provides such a means.  
This basis is based upon the observation that when we decompose
the top quark spin along the direction of the $d$-type quark,
the spin down contribution is small.
The \twotwo\ process produces no spin down $t$ quarks in this
basis, which is equivalent to measuring the helicity of the $t$
quark in the frame where the $t$ quark and $d$-type quark
are back-to-back.  
The overlap contribution, being just the \twotwo\ process
computed with Eq.~(\ref{b-overlap}) in place of the $b$ quark
distribution function, shares a common spin structure
with the \twotwo\ process in this basis.
The amplitude for spin-down $t$ quark
production via the \twothree\ diagrams is suppressed by its
lack of a singularity when the $b$ quark mass is taken to 
zero\cite{TeV1top}.   Since the $d$-type quark appears in
the spectator jet 80\% of the time, if we simply use the
direction of the spectator jet as the top quark spin axis,
we obtain a high degree of polarization:  92\% of the
top quarks associated with the \twothree\ process are
produced with spin up in this basis.  Combining the three
contributions, we find that over-all fraction of spin up
quarks in the spectator basis is 95\%.


For $\bar{t}$ production the situation is a bit different.
The $d$-type quark is in the final state only 31\% of
the time;  in the remainder of the events, it is supplied by
one of the beams.  Hence, the spectator basis chooses the
``wrong'' direction for the spin axis the majority of the time!
However, the spectator jet is simply the scattered
light quark.  In the $Wg$ fusion process, the momentum
transfer via the $t$-channel $W$ boson deflects the incoming light
quark just a little.  Thus, the spectator jet momentum points
in nearly the same direction as the original light quark momentum.
This fact is reflected in the large (absolute) values of
pseudorapidity at which the spectator jet usually emerges.
Since the spectator jet and initial light quark posses nearly
parallel
momentum vectors, it does not degrade the degree of spin
polarization 
very much to use the spectator jet direction even when
the $d$-type quark was actually in the initial state.
Overall, we find that 93\% of the $\bar{t}$'s
are produced with spin down in the spectator basis, which is only
slightly worse than the situation for $t$'s.

Since the $d$-type quark really comes from one of the two
beams the majority of the time, it is worthwhile to consider
the beamline basis in addition to the spectator basis.
From Ref.~\cite{TopPairsHadronic} we recall that the beamline
basis is defined by decomposing the $\bar{t}$ spin along the
direction of one of the beams as seen in the $\bar{t}$ rest frame.
Hence, there are two different beamline bases, since the two beams
are not back-to-back in the $\bar{t}$ rest frame.  We want to
choose the beam which supplied the light quark.  
As we noted in the previous paragraph, the spectator jet typically
points in the same direction as the beam which supplied the
light quark.  Therefore, we should choose to decompose the $\bar{t}$
spin along the beam which is most-nearly aligned with the spectator
jet on an event-by-event basis.  That is, we define the
$\eta$-beamline basis as follows: 
if the pseudorapidity of the spectator jet is positive, 
choose the right-moving beam as the spin axis.  
If the pseudorapidity of the spectator jet is negative, 
choose the left-moving beam as the spin axis.  
In terms of the $\eta$-beamline basis we
find that 90\% of the $\bar{t}$'s have spin down.  While this
is somewhat worse than simply using the spectator basis, 
matters may be improved by using only those events where the
spectator jet has a pseudorapidity which is larger in magnitude
than some cut value $\eta_{min}$.  This takes advantage of the
fact that an initial state $d$ quark is a valence quark.
Thus, on average, it carries a bigger longitudinal momentum
fraction than a quark plucked from the sea.  As a result,
the spectator jet from such events tends to be produced at (slightly)
larger pseudorapidity than events initiated by $\bar{u}$, 
$s$, or $\bar{c}$ quarks.  
So a minimum pseudorapidity requirement increases the
chances that the chosen beam actually does contain the $d$-type quark.
For example, choosing $\eta_{min} = 2.5$ results in a spin
decomposition very similar to that obtained in the spectator basis.
Because such a cut on the spectator jet pseudorapidity is
envisioned by the experiments in order to separate the signal from
the background\cite{ATLAS-report}, 
there is no disadvantage to including a minimum
$\vert\eta\vert$ requirement in our definition of the $\eta$-beamline
basis.

For convenience, we have summarized our results
for the spin decompositions of single $t$ and $\bar{t}$
production at the LHC in
Tables~\ref{Tspin} and~\ref{TBARspin}.
The final column of both tables contains the spin asymmetry
$(N_\uparrow - N_\downarrow)/(N_\uparrow + N_\downarrow)$.
This is the quantity which appears in the differential
distribution of the decay angle~\cite{Spin}:
\beq
{{1}\over{\sigma_T}}\ts
{{d\sigma}\over{d\cos\theta}} =
{{1}\over{2}} \ts
\Biggl[
1 + 
{  {N_\uparrow-N_\downarrow}\over{N_\uparrow+N_\downarrow} }
\cos\theta
\Biggr].
\label{Observable}
\eeq
In Eq.~(\ref{Observable}), $\theta$ is the angle between the charged
lepton (from the decaying top quark) and the chosen spin axis,
as measured in the top quark rest frame.\footnote{To describe
the decay of a $\bar{t}$ quark, we should replace
$\cos\theta$ by $-\cos\theta$ in Eq.~(\protect\ref{Observable}).
Since the $\bar{t}$'s are primarily spin down in the bases we
are considering, the $\bar{t}$ spin asymmetry will be negative.
Thus, the $t$ and $\bar{t}$ samples
may be combined without diluting the resulting angular correlations
in the event that the sign of the charged lepton cannot be
determined.}
Obviously, we want to make the spin asymmetry as large as possible
in order to make this angular correlation easier to observe.
From Tables~\ref{Tspin} and~\ref{TBARspin} we see that the
spectator basis produces correlations which are about a factor
of 3 larger than in the LAB helicity basis.
The improvement provided by the $\eta$-beamline basis is
comparable.

\begin{figure}[h]

\vspace*{9cm}
\includegraphics{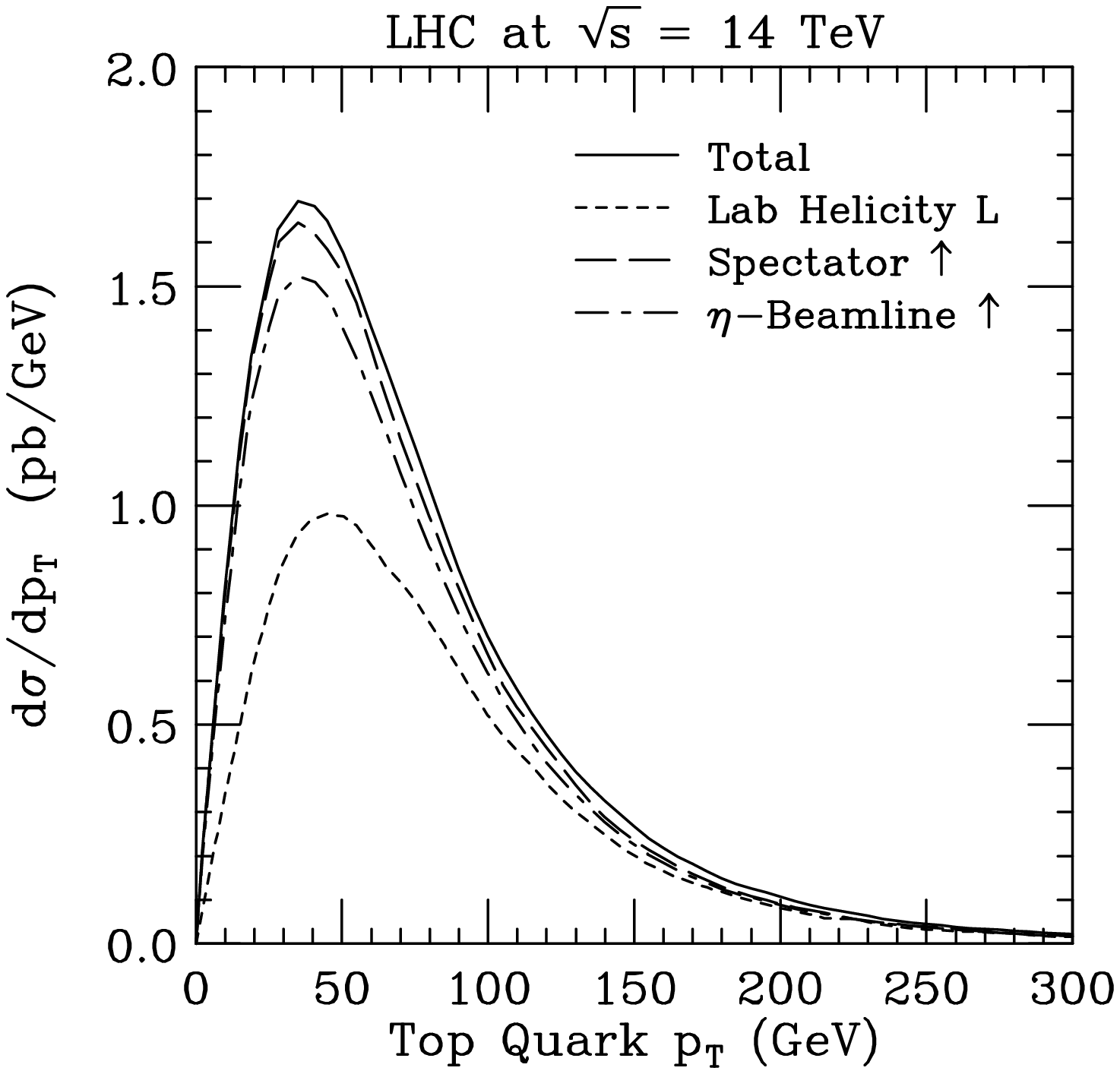}

\vspace{2.0cm}
\caption[]{The differential cross sections (total, LAB helicity
basis left, spectator basis up, and $\eta$-beamline basis up
(with $\eta_{min}=0$)) as a function of the top quark transverse 
momentum for single top quark production via $Wg$ fusion at the
LHC with a center of mass energy of 14 TeV.
}
\label{pTfig}
\end{figure}

\begin{figure}[h]

\vspace*{9cm}
\includegraphics{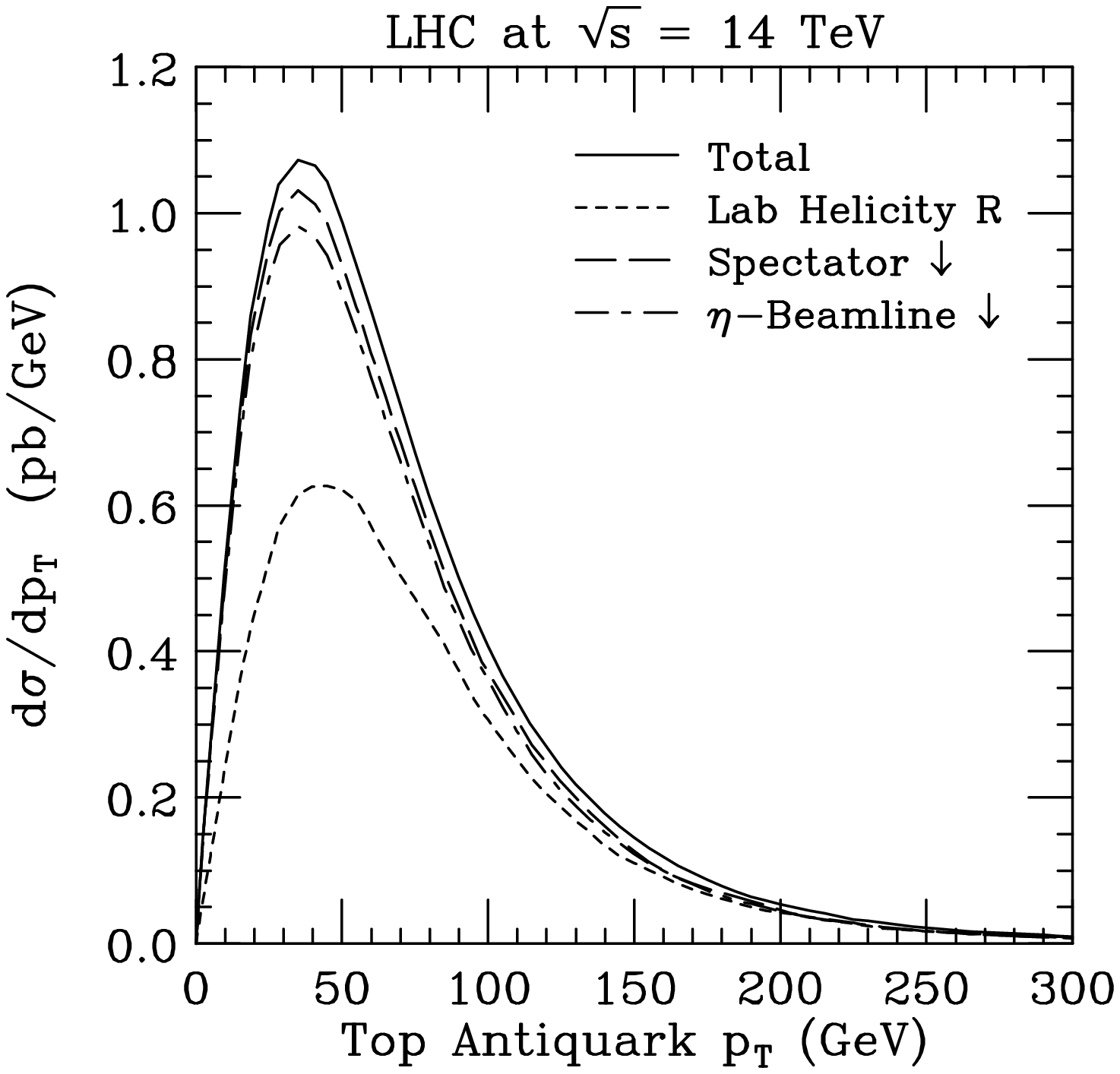}

\vspace{2.0cm}
\caption[]{The differential cross sections (total, LAB helicity
basis right, spectator basis down, and $\eta$-beamline basis down
(with $\eta_{min}=0$)) as a function of the top antiquark transverse 
momentum for single top antiquark production via $Wg$ fusion at the
LHC with a center of mass energy of 14 TeV.
}
\label{pTfig2}
\end{figure}

Figs.~\ref{pTfig} and~\ref{pTfig2} present the $p_T$
distributions of the produced $t$ and $\bar{t}$ quarks.
In addition to the total cross section, we have plotted 
the contributions from the dominant spin component in
the LAB helicity, spectator, and $\eta$-beamline bases.


In general, the spin of the top quark depends upon the point in
phase space at which it is produced.  Therefore, it is important
to make an assessment of the impact of the experimental cuts
which are imposed to isolate the signal from the background.
Although a full-scale detector simulation is beyond the scope
of this letter, we have investigated the effect of the following
theorists' cuts:
\beqa
\begin{array}{lcl}
\hbox{missing energy:}  &\phantom{\Bigl[}\quad&   
                         \not{\nts p\ts}_T>15\GeV,  \\
\hbox{lepton:}          &\phantom{\Bigl[}\quad&   
                         p_T>15\GeV, \ns\vert\eta\vert<2.5 \\
\hbox{spectator jet:}   &\phantom{\Bigl[}\quad&   
                         p_T>50\GeV, 
                         \ns 2.5 < \vert\eta\vert < 5.0 \\
\hbox{bottom jet:}      &\phantom{\Bigl[}\quad&   
                         p_T>50\GeV, \ns\vert\eta\vert<2.5 \\
\hbox{isolation cut:}   &\phantom{\Bigl[}\quad& 
                         \sqrt{(\Delta\eta)^2 + 
                               (\Delta\varphi)^2} > 0.4,
                         \ns\hbox{all pairs} \\
\hbox{third jet:}       &\phantom{\Bigl[}\quad&   
                          \hbox{none with}\ns p_T>30\GeV, 
                          \ns\vert\eta\vert<2.5 .
\end{array}
\label{CUTS}
\eeqa
These cuts are similar to the ones used in the ATLAS 
design study~\cite{ATLAS-report}.
Because these cuts tend to bias towards events where the top
quark has a large velocity in the ZMF (to the extent that the ZMF
can be defined), we expect that the spin fractions will be higher
in their presence.  Indeed this is the case, as may be seen from
Tables~\ref{TspinCUT} and~\ref{TBARspinCUT}.
Even in the presence of cuts, both the spectator and $\eta$-beamline
bases outperform the LAB helicity basis by more than a factor
of 2 with regard to the magnitude of the angular correlations
present in the final state.  For $t$ events, the spectator basis
is slightly better than the $\eta$-beamline basis.  For $\bar{t}$
events, the $\eta$-beamline basis is slightly better than the
spectator basis.  In both cases, however, the differences are so
small that it would certainly be worthwhile to do the experimental
analysis using both bases, especially since some of the systematics
will differ in the two cases.


To summarize, we have seen that it is not possible to uniquely
define the zero momentum frame of the initial state partons in
the $Wg$-fusion process at the LHC.  Consequently, when studying
the spin of the produced top quarks, it does not make sense to
use the ZMF helicity basis.  Instead, we must use a spin basis
whose definition does not depend on the existence of a well-defined
ZMF.    Simply using the LAB helicity basis results in a
description of the top quarks where both spin components are
comparable in size.  
However, there are spin bases where the top quarks are described
primarily by just one of the two possible spin states.
Two such bases are the spectator and $\eta$-beamline bases.
In the spectator basis, we decompose the spin of the top quark 
in its rest frame along the direction of the spectator jet 
as seen in that frame.   In the $\eta$-beamline basis we decompose
the spin of the top quark in its rest frame along the direction
of one of the proton beams as seen in that frame.  The right-moving
beam is chosen if the pseudorapidity of the spectator jet is
positive, whereas the left-moving beam is chosen if the pseudorapidity
of the spectator jet is negative.  We find that both of these bases
the spin angular correlations are approximately a factor of 3 larger
than in the LAB helicity basis.  The utility of these two
bases is not adversely affected by the imposition of the sorts of
cuts required to extract a $Wg$ fusion signal from the background.


\acknowledgements

We would like to thank Dugan O'Neil for prompting
us to think carefully about the relationship
between the $2\rightarrow2$ and $2\rightarrow3$ processes.
We would also like to thank Scott Willenbrock and Zack Sullivan
for their helpful discussions concerning
their calculation\cite{Stelzer1}
of the next-to-leading order $Wg$-fusion cross section.
SJP would like to thank the SLAC theory group
for their support and hospitality during the initial stages of 
this work.

The Fermi National Accelerator Laboratory is operated by 
Universities Research Association, Inc., under contract 
DE-AC02-76CHO3000 with the U.S.A. Department of Energy. 
High energy physics research at McGill University is
supported in part by the Natural Sciences and Engineering 
Research Council of Canada and the Fonds pour la Formation de 
Chercheurs et l'Aide \`a la Recherche of Qu\'ebec.
SLAC is supported by the U.S.A. Department of Energy under contract
DE-AC03-76SF00515.



\begin{table}
\caption{Fractional cross sections for single top and single
antitop production in
the $Wg$ fusion channel at the LHC at 14.0 TeV, decomposed
according to the flavor of the light quark appearing in
the initial state.
\label{Flavors}}
\begin{tabular}{ccddc}
& $q$ &  $t$ & $\bar{t}$ & \\
\hline
& $u$       & 74\% & 20\% & \\
& $d$       & 12\% & 56\% & \\
& $s$       & \ns 8\% & 13\% & \\
& $c$       & \ns 6\% & 11\% &
\end{tabular}
\end{table}

\begin{table}
\caption{Dominant spin fractions and asymmetries for the various
bases studied for single top quark production in the $Wg$ 
fusion channel at the LHC at 14.0 TeV.  
In addition to the total spin fractions, 
we have listed the fractions associated
with each of the three types of diagrams 
(Figs.~\protect\ref{WgFusionDiagrams}a--c) contributing to the
total.
\label{Tspin}}
\begin{tabular}{cccccd}
basis & \twotwo\ & overlap & \twothree\ & total &
$\displaystyle{{{N_{\up}-N_{\down}}
\over{N_{\up}+N_{\down}}}}$  \\
\hline
LAB helicity &  66\% $\downarrow$(L) 
             &  64\% $\downarrow$(L)
             &  59\% $\downarrow$(L)
             &  64\% $\downarrow$(L)
             &  $-$0.27               \\
ZMF helicity &  99\% $\downarrow$(L) 
             &  undefined
             &  82\% $\downarrow$(L)
             &  undefined
             &  undefined             \\
spectator    &  99\% $\uparrow$\phantom{(L)}
             &  99\% $\uparrow$\phantom{(L)}
             &  92\% $\uparrow$\phantom{(L)}
             &  95\% $\uparrow$\phantom{(L)}
             &  0.89                  \\
$\eta$-bml, $\eta_{min}=0$   
             &  93\% $\uparrow$\phantom{(L)}
             &  93\% $\uparrow$\phantom{(L)}
             &  86\% $\uparrow$\phantom{(L)}
             &  88\% $\uparrow$\phantom{(L)}
             &  0.77                  \\
$\eta$-bml, $\eta_{min}=2.5$   
             &  97\% $\uparrow$\phantom{(L)}
             &  97\% $\uparrow$\phantom{(L)}
             &  90\% $\uparrow$\phantom{(L)}
             &  93\% $\uparrow$\phantom{(L)}
             &  0.85                  
\end{tabular}
\end{table}

\begin{table}
\caption{Dominant spin fractions and asymmetries for the various
bases studied for single top antiquark production in the $Wg$ 
fusion channel at the LHC at 14.0 TeV.  
In addition to the total spin fractions, 
we have listed the fractions associated with
each of the three types of diagrams 
(Figs.~\protect\ref{WgFusionDiagrams}a--c) contributing to the
total.
\label{TBARspin}}
\begin{tabular}{cccccd}
basis & \twotwo\ & overlap  & \twothree\ & total &
$\displaystyle{{{N_{\up}-N_{\down}}
\over{N_{\up}+N_{\down}}}}$  \\
\hline
LAB helicity &  67\% $\uparrow$(R) 
             &  64\% $\uparrow$(R)
             &  60\% $\uparrow$(R)
             &  65\% $\uparrow$(R)
             &  0.29                  \\
ZMF helicity &  97\% $\uparrow$(R) 
             &  undefined
             &  78\% $\uparrow$(R)
             &  undefined
             &  undefined             \\
spectator    &  98\% $\downarrow$\phantom{(R)}
             &  97\% $\downarrow$\phantom{(R)}
             &  91\% $\downarrow$\phantom{(R)}
             &  93\% $\downarrow$\phantom{(R)}
             &  $-$0.87                \\
$\eta$-bml, $\eta_{min}=0$
             &  94\% $\downarrow$\phantom{(R)}
             &  94\% $\downarrow$\phantom{(R)}
             &  88\% $\downarrow$\phantom{(R)}
             &  90\% $\downarrow$\phantom{(R)}
             &  $-$0.79                 \\
$\eta$-bml, $\eta_{min}=2.5$
             &  99\% $\downarrow$\phantom{(R)}
             &  99\% $\downarrow$\phantom{(R)}
             &  91\% $\downarrow$\phantom{(R)}
             &  94\% $\downarrow$\phantom{(R)}
             &  $-$0.87
\end{tabular}
\end{table}

\begin{table}
\caption{Dominant spin fractions and asymmetries for the various
bases studied for single top quark production in the $Wg$ 
fusion channel at the LHC at 14.0 TeV, subject to the set of
cuts described in Eq.~(\protect\ref{CUTS}).
In addition to the total spin fractions, 
we have listed the fractions associated
with each of the three types of diagrams 
(Figs.~\protect\ref{WgFusionDiagrams}a--c) contributing to the
total.
\label{TspinCUT}}
\begin{tabular}{cccccd}
basis & \twotwo\ & overlap & \twothree\ & total &
$\displaystyle{{{N_{\up}-N_{\down}}
\over{N_{\up}+N_{\down}}}}$  \\
\hline
LAB helicity & \phantom{1}75\% $\downarrow$(L) 
             & \phantom{1}74\% $\downarrow$(L)
             &  71\% $\downarrow$(L)
             &  74\% $\downarrow$(L)
             &  $-$0.48               \\
spectator    & 100\% $\uparrow$\phantom{(L)}
             & 100\% $\uparrow$\phantom{(L)}
             &  99\% $\uparrow$\phantom{(L)}
             &  99\% $\uparrow$\phantom{(L)}
             &  0.99                  \\
$\eta$-bml   & \phantom{1}99\% $\uparrow$\phantom{(L)}
             & \phantom{1}99\% $\uparrow$\phantom{(L)}
             &  98\% $\uparrow$\phantom{(L)}
             &  98\% $\uparrow$\phantom{(L)}
             &  0.96                  
\end{tabular}
\end{table}

\begin{table}
\caption{Dominant spin fractions and asymmetries for the various
bases studied for single top antiquark production in the $Wg$ 
fusion channel at the LHC at 14.0 TeV, subject to the set of
cuts described in Eq.~(\protect\ref{CUTS}).
In addition to the total spin fractions, 
we have listed the fractions associated
with each of the three types of diagrams 
(Figs.~\protect\ref{WgFusionDiagrams}a--c) contributing to the
total.
\label{TBARspinCUT}}
\begin{tabular}{cccccd}
basis & \twotwo\ & overlap  & \twothree\ & total &
$\displaystyle{{{N_{\up}-N_{\down}}
\over{N_{\up}+N_{\down}}}}$  \\
\hline
LAB helicity & \phantom{1}72\% $\uparrow$(R) 
             & \phantom{1}70\% $\uparrow$(R)
             &  67\% $\uparrow$(R)
             &  70\% $\uparrow$(R)
             &  0.41                  \\
spectator    & \phantom{1}99\% $\downarrow$\phantom{(R)}
             & \phantom{1}99\% $\downarrow$\phantom{(R)}
             &  97\% $\downarrow$\phantom{(R)}
             &  98\% $\downarrow$\phantom{(R)}
             &  $-$0.96                \\
$\eta$-bml   & 100\% $\downarrow$\phantom{(R)}
             & 100\% $\downarrow$\phantom{(R)}
             &  98\% $\downarrow$\phantom{(R)}
             &  99\% $\downarrow$\phantom{(R)}
             &  $-$0.97
\end{tabular}
\end{table}

\end{document}